%
%
%

\magnification=1200
\baselineskip=14pt
\def\d{{\rm d}}
\def\o{\omega}
\def\O{\Omega}
\def\l{\lambda}
\def\ln{{\rm log\,}}
\def\D{\partial}

\def\G{\Gamma}
\def\F{{\cal F}}
\def\R{{\rm Res}}
\def\a{\tilde a}
\def\L{\bar\Lambda}
\def\tD{\tilde\partial}
\def\S{\tilde S}
\def\+{{\rm Res}_{P_+}(z\d\l)}
\def\-{{\rm Res}_{P_-}(z\d\l)}

\rightline{UCLA/96/TEP/40}
\rightline{Columbia/Math/96}
\bigskip
\centerline{\bf THE RENORMALIZATION GROUP EQUATION IN N=2}
\bigskip
\centerline{{\bf SUPERSYMMETRIC GAUGE THEORIES}\footnote*{Research
supported in part by the National Science Foundation
under grants PHY-95-31023 and DMS-95-05399}}
\bigskip
\bigskip
\centerline{{\bf Eric D'Hoker}${}^1$, 
												{\bf I.M. Krichever}${}^2$,
            {\bf and D.H. Phong} ${}^3$}
\bigskip
\centerline{${}^1$ Department of Physics}
\centerline{University of California, Los Angeles, CA 90024, USA}
\centerline{e-mail: dhoker@physics.ucla.edu}
\bigskip 
\centerline{${}^2$ Landau Institute for Theoretical Physics}
\centerline{Moscow 117940, Russia}
\centerline{and} 
\centerline{Department of Mathematics}
\centerline{Columbia University, New York, NY 10027, USA}
\centerline{e-mail: krichev@math.columbia.edu}
\bigskip
\centerline{${}^3$ Department of Mathematics}
\centerline{Columbia University, New York, NY 10027, USA}
\centerline{e-mail: phong@math.columbia.edu}
\bigskip
\centerline{\bf Abstract}
\medskip
We clarify the mass dependence of the effective prepotential
in N=2 supersymmetric SU($N_c$) gauge theories with an arbitrary
number $N_f<2N_c$ of flavors. The resulting
differential equation for the prepotential extends the equations
obtained previously for SU(2) and for zero masses. It can
be viewed as an exact renormalization group equation for the
prepotential, with the beta function given by a modular form.
We derive an explicit formula for this modular form when $N_f=0$,
and verify the equation to 2-instanton order in the weak-coupling
regime for arbitrary $N_f$ and $N_c$.

\vfill\break

\centerline{\bf I. INTRODUCTION}
\bigskip
New avenues for the investigation 
of N=2 supersymmetric
gauge theories have recently opened up
with the Seiberg-Witten proposal [1], which
gives the effective action
in terms of a 1-form $\d\l$ on 
Riemann surfaces fibering
over the moduli space of vacua.
Starting with the SU(2) theory [1], a form $\d\l$
is now available for many
other gauge groups [2], with
matter in the fundamental [3][4] or 
in the adjoint representation [5].
This has led to a wealth of information about the
prepotential, including its expansion up to
2-instanton order for asymptotically free theories with
classical gauge groups [6].
\medskip
These developments suggest a rich structure for the
prepotential $\F$, which
may help understand its strong coupling behavior, and
clarify its relation with the point particle limit
of string theories, when gravity is turned off [7].
Of particular interest in this context are the
non-perturbative differential equations derived by Matone in [8] for SU(2),
and later extended by Eguchi and Yang in [9] to SU($N_c$) theories
with only massless matter.
It was however unclear how these equations
would be affected if the hypermultiplets acquire
non-vanishing masses.
\medskip
In the present paper, we address this issue by
providing a systematic and
general framework for incorporating arbitrary masses $m_j$. 
In effect, the masses $m_j$ are treated on an equal 
footing as the vev's $a_k$ of the scalar field in 
the chiral multiplet, since they are both given by 
periods of $\d\l$ around
non-trivial cycles. For the masses, the cycles are small loops
around the poles of $\d\l$, while for $a_k$, they are
non-trivial $A$-homology cycles.
This suggests that the derivatives of $\F$ with respect to
the masses should be given by the periods of $\d\l$ around
"dual cycles", just as the derivatives of $\F$ with respect
to $a_k$ are given by the periods of $\d\l$ around
$B$-cycles. We provide an explicit
closed formula for such a prepotential,
motivated by 
the $\tau$-function of the Whitham hierarchy 
obtained in [10].
(In this connection, we should point out that 
intriguing similarities between
supersymmetric gauge theories and Whitham hierarchies
had been noted by many authors [11],
and had been the basis of the considerations in [9],
as well as in [4], the starting point of our
arguments). Written in terms of the derivatives of $\F$,
this closed formula becomes the non-perturbative
equation for $\F$ that we seek.
It can be verified explicitly to 2-instanton order,
using the results of [6].
\medskip
Specifically, the differential equation for $\F$ is of the form 
$${\cal D}\F=
-{1\over 2\pi i}\big[\R_{P_-}(z\d\l)\R_{P_-}(z^{-1}\d\l)
+\R_{P_+}(z\d\l)\R_{P_+}(z^{-1}\d\l)\big]\eqno(1.1)
$$
with ${\cal D}$ the operator 
$$
{\cal D}
=\sum_{k=1}^{N_c}a_k{\D\over\D a_k}
+\sum_{j=1}^{N_f}m_j{\D\over\D m_j}-2
\eqno(1.2)
$$ 
The right hand side in (1.1) has been interpreted 
in [8][9] in terms of the trace of the classical vacuum expectation value 
$\sum_{k=1}^{N_c}\bar a_k^2$, although there are ambiguities
with this interpretation when $N_f\geq N_c$.
Mathematically, it can be expressed in terms
of $\vartheta$-functions for arbitrary $N_c$
when $N_f=0$ (c.f. Section III (c) below). 
There is little doubt that
this should be the case in general.
Now we have by dimensional analysis 
$$
({\cal D}
+\Lambda{\D\over\D\Lambda})\F=0\eqno(1.3)
$$ 
if $\Lambda$ is the
renormalization scale of the theory.
Thus the proper interpretation for
the equation (1.1) is as
a renormalization group equation, with the
beta function given by a modular form!
\medskip 
Finally, we observe that the
effective Lagrangian in the low momentum
expansion
determines the effective prepotential only
up to $a_k$-independent terms. However, masses
can arise as vacuum expectation values
of non-dynamical fields, and we would expect
the natural dependence on masses imposed here to
be useful in future developments, for example
in eventual generalizations to string theories.

\bigskip
\bigskip

\centerline{\bf II. A CLOSED FORM FOR THE PREPOTENTIAL}

\bigskip

\noindent{\bf 
(a) The geometric set-up for N=2 supersymmetric gauge theories}

We recall the basic set-up for the effective prepotential $\F$ of 
N=2 supersymmetric SU($N_c$)
gauge theories.
\medskip
The moduli space of vacua is an $N_c-1$ dimensional variety,
which can be parametrized classically by the eigenvalues
$\bar a_k$, $\sum_{k=1}^{N_c}\bar a_k=0$
of the scalar field $\phi$ in the adjoint representation
occurring in the N=2 chiral multiplet. (The flatness of the potential
is equivalent to $[\phi,\phi^{\dagger}]=0$). Quantum mechanically,
the order parameters $\bar a_k$ get renormalized to 
parameters $a_k$. The prepotential $\F$ determines
completely the Wilson effective Lagrangian
of the quantum theory to leading order in the low momentum
expansion. 
Following Seiberg-Witten [1], we require that the renormalized
order parameters $a_k$, their duals $a_{D,k}$,
and the prepotential $\F$ be given by
$$
\eqalignno{
 a_k&={1\over 2\pi i}\oint_{A_k}\d\l,
\qquad
\ a_{D,k}={1\over 2\pi i}\oint_{B_k}\d\l \cr
{\D\F\over\D a_k}&=a_{D,k}&(2.1)\cr}
$$ 
where $\d\l$ is a suitably chosen meromorphic 1-form on a fibration
of Riemann surfaces $\G$ above the moduli space of vacua,
and $A_j$, $B_j$ is a canonical basis of homology cycles on $\G$.
\medskip
In the formalism of [4], the form $\d\l$ 
is characterized by two meromorphic Abelian differentials $\d Q$ and
$\d E$ on $\G$, with $\d\l=Q\d E$. For SU($N_c$) gauge theories
with $N_f$ hypermultiplets in 
the fundamental representation, $N_f<2N_c$,
the defining properties of $\d E$ and $\d Q$ are

\item{$\bullet$} $\d E$ has only simple poles, at points $P_+$, $P_-$,
$P_i$, where its residues are
respectively $-N_c$, $N_c-N_f$, and $1$ ($1\leq i\leq N_f$).
Its periods around homology cycles are integer multiples of $2\pi i$;

\item{$\bullet$} $Q$ is a well-defined meromorphic $function$,
which has simple poles at $P_+$ and $P_-$, and takes the values
$Q(P_i)=-m_i$ at $P_i$, where $m_i$ are the bare masses
of the $N_f$ hypermultiplets;

\item{$\bullet$} The form $\d\l$ is normalized so that
$$
\eqalignno{\ {\rm Res}_{P_+}(z\d\l)&=-N_c2^{-1/N_c},
\ {\rm Res}_{P_-}(z\d\l)=(N_c-N_f) 
({\Lambda^{2N_c-N_f}\over 2})^{1/(N_c-N_f)}\cr
{\rm Res}_{P_+}(\d\l)&=0&(2.2)\cr}
$$
where $\Lambda$ is the dynamically generated scale of the theory,
and $z=E^{-1/N_c}$ or $z=E^{1/(N_c-N_f)}$ is the holomorphic 
coordinate system provided by the Abelian integral $E$, depending
on whether we are near $P_+$ or near $P_-$.
\medskip
It was shown in [4] that these conditions imply that $\G$ is hyperelliptic,
and admits an equation of the form
$$
y^2=\big(\prod_{k=1}^{N_c}(Q-\tilde a_k)\big)^2-\Lambda^{2N_c-N_f}
\prod_{j=1}^{N_f}(Q+m_j)\equiv A(Q)^2-B(Q)\eqno(2.3)
$$ 
Here $\tilde a_k$ are parameters which coincide with $\bar a_k$
when $N_c<N_f$, but may otherwise receive corrections. It is convenient
to set
$$
\bar\Lambda=\Lambda^{{1\over 2}(2N_c-N_f)}
$$
The function $Q$ in $\d\l=Q\d E$ is now the coordinate $Q$ 
in the complex plane,
lifted to the two sheets $y=\pm\sqrt{A^2-B}$ of (2.3), while
the Abelian integral $E$ is given by $E=\ln (y+A(Q))$. 
The points $P_{\pm}$
correspond to $Q=\infty$, with the choice of signs $y=\pm\sqrt{A^2-B}$. 
\medskip
\noindent{\bf (b) The prepotential in closed form}

We shall now exhibit a solution $\F$ for the equations (2.1)
in closed form. Formally, it is given by
$$
\eqalignno{
2\F={1\over 2\pi i}&
\bigg[\sum_{k=1}^{N_c}a_k\oint_{B_k}\d\l
-\sum_{j=1}^{N_f}m_j\int_{P_-}^{P_j}\d\l\cr
&+\+\R_{P_+}(z^{-1}\d\l)
+\-
\R_{P_-}(z^{-1}\d\l)\bigg]
&(2.4)\cr}
$$
However, the above expression involves divergent
integrals which must be regularized. For this, 
we need to make a number of choices. First,
we fix a canonical homology basis $A_i$, $B_i$, along which the
Riemann surface can be cut out to obtain a domain with
boundary $\prod_{i=1}^{N_c-1}A_i^{-1}B_i^{-1}A_iB_i$. 
Next, we fix simple paths
$C_-$, $C_j$ from $P_+$ to $P_-$, $P_j$ respectively
($1\leq j\leq N_f$), which have only $P_+$ as common point.
As usual the cuts are viewed as having two edges.
With these choices, we can define a single-valued branch
of the Abelian integral $E$ in 
$\G_{cut}=\G\setminus(C_-\cup C_1\cup\cdots\cup C_{N_f})$
as follows. Near $P_+$, the function $Q^{-1}$ provides a biholomorphism
of a neighborhood of $P_+$ to a small disk in the complex plane.
Choose the branch of $\ln Q^{-1}$ with a cut along $Q^{-1}(C_-)$,
and define an integral $E$ of $\d E$ in a neighborhood of $P_+$ in
$\G_{cut}$ by requiring that
$$
E=N_c\ln Q+\ln 2+O(Q^{-1})\eqno(2.5)$$
The Abelian integral $E$ can then uniquely defined on $\G_{cut}$ by
integrating along paths. It determines in turn a coordinate
system $z$ near each of the poles 
$P_+$, $P_-$, and $P_j$, $1\leq j\leq N_f$, e.g.,
$$
z=e^{-{1\over N_c}E}\ \ {\rm near}\ \ P_+\eqno(2.6)
$$
It is easily seen that $z$ is holomorphic around $P_+$, and that
$z=2^{1\over N_c}Q^{-1}+O(Q^{-2})$. The next few terms of the expansion
of $z$ in terms of $Q^{-1}$ are actually quite important,
but we shall evaluate them later. Similarly,
we set $z=e^{{1\over N_c-N_f}E}$ near $P_-$, and $
z=e^{-E}$ near $P_j$, $1\leq j\leq N_f$.

The same choices above allow us to define at the same time a 
single-valued
branch of the Abelian integral $\l$ in $\G_{cut}$. Specifically,
$\l$ is defined near $P_+$ by the normalization
$$
\l(z)=-\+{1\over z}+O(z)\eqno(2.7)$$
with $z$ the above holomorphic coordinate (2.6). As before,
$\l$ is then extended to the whole of $\G_{cut}$ by
analytic continuation. Evidently, near $P_-$, $\l$ can be expressed as
$$
\l(z)=-\-{1\over z}+
\l(P_-)+O(z)\eqno(2.8)
$$
in the corresponding coordinate $z$ near $P_-$,
for a suitable constant $\l(P_-)$. Similarly, near $P_j$,
$\l$ can be expressed as
$$
\l(z)=-m_j\ln z+\l(P_j)+O(z)\eqno(2.9)
$$
for suitable constants $P_j$. The expression (2.4)
for the prepotential $\F$ can now be given a precise
meaning by regularizing as follows the divergent integrals
appearing there
$$
\int_{P_-}^{P_j}\d\l=\l(P_j)-\l(P_-)\eqno(2.10)
$$
This method of regularization has the advantage
of commuting with differentiation under
the integral sign with respect to
connections which keep the values of $z$ constant.
\medskip
\noindent{\bf (c) The derivatives of the prepotential}

The main properties of $\F$ are the following
$$
\eqalignno{
{\D\F\over\D a_k}&={1\over 2\pi i}\oint_{B_k}\d\l&(2.11)\cr
{\D\F\over\D m_j}&={1\over 2\pi i}\bigg[-\int_{P_-}^{P_j}\d\l
+{1\over 2}\sum_{i=1}^{N_f}m_i\big(\int_{P_-}^{P_i}\d\O_j^{(3)}
-\int_{P_-}^{P_j}\d\O_i^{(3)}\big)\bigg]&(2.12)\cr}
$$
where $\d\O_i^{(3)}$ are Abelian differentials of the third
kind with simple poles and residues +1 and -1 at $P_-$ and $P_i$
respectively, normalized to have vanishing
$A_j$-periods. We observe that the Wilson effective action
of the gauge theory is insensitive to
modifications of $\F$ by $a_k$-independent terms.
The equation (2.12) can be viewed as an additional
criterion for selecting $\F$, motivated by the fact that
the mass parameter $-m_j$ of $\d\l$ can be viewed as a contour integral
of $\d\l$ around a cycle surrounding the pole $P_j$.
In analogy with (2.4), the derivatives with respect to $m_j$
of a natural choice for $\F$
should then reproduce
the integral of $\d\l$ around a dual cycle. This is the origin
of the first term on
the right hand side of (2.12), if we view the
path from $P_-$ to $P_j$ as such a dual "cycle".
The second term on the right hand side of (2.12) is
a harmless correction due to regularization. The expression
between parentheses is actually 
always a multiple of
$\pi i$, although we do not need this fact.
\medskip
We now establish (2.11) and (2.12). We need to consider the derivatives
of $\d\l$ with respect to both $a_k$ and $m_j$.
We use the connection $\nabla^E=\nabla$ of [4], which differentiates
along subvarieties where the value of the Abelian integral
$E$ (equivalently the coordinate $z$) is kept constant.
Then simply by inspecting the derivatives of the singular
parts of $\d\l$ in a Laurent expansion in the $z$-coordinate
near each pole, we find that
$$
\nabla_{a_k}\d\l=2\pi i\,\d\omega_k,\ \nabla_{m_j}\d\l=\d\Omega_j^{(3)},
\eqno(2.13)
$$
where $\d\o_k$ is the basis of Abelian differentials of the
first kind dual to the $A_k$-cycles. Next, we recall from
(2.2) that the residues $\+$ and $\-$ are constant. Consequently,
$$
\eqalignno{2{\D\F\over\D a_k}=
{1\over 2\pi i}&
\oint_{B_k}\d\l+\sum_{i=1}^{N_c}a_i
\oint_{B_i}\d\omega_k
-\sum_{j=1}^{N_f}m_j\int_{P_-}^{P_j}\d\omega_k\cr
&+\+\R_{P_+}(z^{-1}
\d\omega_k)
+\-
\R_{P_-}(z^{-1}\d\omega_k)&(2.14)\cr}
$$
However, we also have the following Riemann bilinear relations,
valid even in presence of regularizations
$$
\eqalignno
{\oint_{B_i}\d\o_k&=\oint_{B_k}\d\o_i, \cr 
{1\over 2\pi i}\oint_{B_k}\d\Omega_j^{(3)} & =-\int_{P_-}^{P_j}\d\o_k\cr
{1\over 2\pi i}\oint_{B_k}\d\O_{\pm}^{(2)}&=\R_{P_{\pm}}(z^{-1}\d\o_k)\cr
\int_{P_-}^{P_j}\d\O_{\pm}^{(2)}&=-\R_{P_{\pm}}
(z^{-1}\d\O_j^{(3)})&(2.15)\cr}
$$
Here
$\d\Omega_{\pm}^{(2)}$ are Abelian differentials of
the second kind, with a double pole
at $P_{\pm}$, vanishing $A$-cycles,
and normalization
$$
\d \Omega_{\pm}^{(2)}=z^{-2}\d z+O(z)\eqno(2.16)
$$
The relations (2.15) follow from the usual Riemann bilinear arguments,
by considering respectively the (vanishing) integrals on the cut
surface $\G_{cut}$ of
the 2-forms $\d(\o_i\d\omega_k)$, $\d(\Omega_j^{(3)}\d\o_k)$,
$\d(\O_{\pm} ^{(2)}\d\o_k)$, $\d(\O_j^{(3)}\d\O_{\pm})$.
Applying (2.15) to (2.14),
we obtain
$$
\eqalignno{2{\D\F\over\D a_k}=&
{1\over 2\pi i}
\oint_{B_k}\d\l+\sum_{i=1}^{N_c}a_i\oint_{B_k}\d\omega_i
+{1\over 2\pi i}\sum_{j=1}^{N_f}m_j\oint_{B_k}\d\Omega_j^{(3)}\cr
&\quad+{1\over 2\pi i}\+
\oint_{B_k}\d\O_+^{(2)}
+{1\over 2\pi i}\-
\oint_{B_k}\d\O_-^{(2)}&(2.17)\cr}
$$ 
However, the expression
$$
\d\l=2\pi i\sum_{i=1}^{N_c}a_i\d\o_i
+\+\d\O_+^{(2)}+
\-
\d\O_-^{(2)}+
\sum_{j=1}^{N_f}m_j\d\O_j^{(3)}
\eqno(2.18)
$$
is just the expansion of $\d\l$ in terms of Abelian differentials
of first, second, and third kind! The equation (2.11) follows.
The equation (2.12) can be established in the same
way. First we write
$$
\eqalignno{2{\D\F\over\D m_l}
=&
{1\over 2\pi i}
\bigg[\sum_{i=1}^{N_c}a_i\oint_{B_i}\d\O_l^{(3)}  
-\int_{P_-}^{P_l}\d\l
-\sum_{j=1}^{N_f}m_j\int_{P_-}^{P_j}\d\O_l^{(3)}\cr
&\quad+\+
\R_{P_+}(z^{-1}\d\O_l^{(3)})
+\-
\R_{P_-}(z^{-1}\d\O_l^{(3)})\bigg]&(2.19)\cr}
$$
Substituting in the bilinear relations gives
$$
\eqalignno{2{\D\F\over\D m_l}
=&
{1\over 2\pi i}
\bigg[-2\pi i\sum_{i=1}^{N_c}a_i\int_{P_-}^{P_l}\d\o_l  
-\int_{P_-}^{P_l}\d\l
-\sum_{j=1}^{N_f}m_j\int_{P_-}^{P_l}\d\O_j^{(3)}\cr
&\quad-\+
\int_{P_-}^{P_l}\d\O_+^{(2)}
-\-
\int_{P_-}^{P_l}\d\O_-^{(2)}\bigg]\cr
&\quad-{1\over 2\pi i}\sum_{j=1}^{N_f}m_j(\int_{P_-}^{P_j}
\d\O_l^{(3)}-\int_{P_-}^{P_l}\d\Omega_j^{(3)})&(2.20)\cr}
$$
Again,
the Abelian differentials recombine to produce $\d\l$,
and the relation (2.12) follows.

\bigskip
\bigskip

\centerline{\bf III. THE RENORMALIZATION GROUP EQUATION}

\bigskip

\noindent{\bf (a) The renormalization group equation in terms 
of residues}

Combining the equations (2.4), (2.11), and (2.12) gives
a first version of the renormalization group equation for $\F$, valid
in presence of arbitrary masses $m_j$
$$
\eqalignno{
\sum_{k=1}^{N_c}a_k{\D\F\over\D a_k}
+\sum_{j=1}^{N_f}m_j{\D\F\over\D m_j}-2\F
=
-{1\over 2\pi i}\bigg[&\+\R_{P_+}(z^{-1}\d\l)\cr
&+\-
\R_{P_-}(z^{-1}\d\l)\bigg]&(3.1)\cr}
$$
\medskip
\noindent{\bf (b) The renormalization group equation 
in terms of invariant polynomials}

We can evaluate the right hand side of (3.1) explicitly,
in terms of the masses $m_j$, and the
moduli parameters $\tilde a_k$ and $\Lambda$ of the spectral
curve (2.3). For this, we need
the first three leading coefficients 
in the expansion of $Q$ in terms of $z$ at $P_+$ and
$P_-$. Now recall that
at $P_+$, $Q\rightarrow\infty$, $y=\sqrt{A^2-B}$, and  
$$
z=(y+A)^{-1/N_c}
\eqno(3.2)
$$
For $N_f<2N_c$, we 
may expand $\sqrt{A^2-B}$ in
powers of $B/A^2$ and write, up to $O(Q^{N_c-3})$
$$
y+A=2\big[A-{1\over 4}{B\over A}-{1\over 16}{B^2\over A^3}\big]
\eqno(3.3)
$$
We consider first the terms in (3.3) of order
up to $O(Q^{N_c-1})$. Then for $N_f\leq 2N_c-2$, only the 
top two terms in $A$ contribute, while for $N_f=2N_c-1$,
we must also incorporate the term $\L^2 x^{N_f-N_c}=\L^2 x^{N_c-1}$
from $B/A$. Thus
$$
A+y=2Q^{N_c}\big[1-(\tilde s_1+\delta_{N_f,2N_c-1}{\L ^2 \over 4})Q^{-1}\big]
+O(Q^{N_c-2})
$$
where we have introduced the notation
$$
\tilde s_i=(-1)^i\sum_{k_1<\cdots<k_i}\tilde a_{k_1}\cdots\tilde a_{k_i},
\ \qquad 
t_i=\sum_{k_1<\cdots<k_i}\tilde t_{k_1}\cdots\tilde t_{k_i}
$$
This leads to the first two coefficients of $z$ in terms of $Q$,
or equivalently, the first two coefficients of $Q$ in terms of $z$
$$
Q=2^{-1/N_c}z^{-1}\big(1+{1\over N_c}(\tilde s_1+
\delta_{N_f,2N_c-1}{\L ^2 \over 4})z\big)
$$
Comparing with (2.2), we see that this confirms the value
of $\R_{P_+}(z\d\l)$ required there, while the condition
that $\R_{P_+}(\d\l)=0$ is equivalent to
$$
\tilde s_1+\delta_{N_f,2N_c-1}{\L ^2 \over 4}=0\eqno(3.4)
$$
Similarly, in the expansion of $A+y$ to order $O(Q^{N_c-2})$,
we must consider separately the cases $N_f<2N_c-2$, $N_f=2N_c-2$,
and $N_f=2N_c-1$, depending on whether the terms
$B/A$ and $B^2/A^3$ contribute to this order.
Taking into account (3.4), we find
$$
Q=2^{-1/N_c}z^{-1}\big(1-{2^{2/N_c}\over N_c}S_2^+z^2\big)+O(z^2)
\eqno(3.5)
$$
with $S_2^+$ defined to be
$$
S_2^+=\tilde s_2-\delta_{N_f,2N_2-2}{\L ^2 \over 4}-\delta_{N_f,2N_c-1}
{\L ^2 \over 4}t_1\eqno(3.6)
$$
Near $P_-$, we have instead 
$$
A+y=A-A(1-{B\over A^2})^{1/2}={1\over 2}{B\over A}
+{1\over 8}{B^2\over A^3}
+{1\over 16}{B^3\over A^5}
$$
Again, considering separately the cases
$N_f<2N_c-2$, $N_f=2N_c-2$, $N_f=2N_c-1$,
we can derive the leading three terms of
the expansion of
$z=E^{-1/(N_f-N_c)}=(A+y)^{-1/(N_f-N_c)}$ in terms
of $Q$. Written in terms of an expansion of $Q$
in terms of $z$, the result is
$$
\eqalignno{Q=&({\L ^2 \over 2})^{-1/(N_f-N_c)}z^{-1}
\bigg[1-{t_1\over N_f-N_c}({\L ^2 \over 2})^{1/(N_f-N_c)}z\cr
&\quad
+{1\over (N_f-N_c)^2}({\L ^2 \over 2})^{2/(N_f-N_c)}\big(S_2^-(N_f-N_c)
+{1\over 2}(-1+N_f-N_c)t_1^2\big)z^2\bigg]
&(3.7)\cr}
$$
with $S_2^-$ given by
$$
S_2^-=\tilde s_2-t_2-\delta_{N_f,2N_2-2}{\L ^2\over 4}-\delta_{N_f,2N_c-1}
{\L ^2 \over 4}t_1\eqno(3.8)
$$
Since $\d\l=-N_cQ{\d z\over z}$ near $P_+$ and 
$\d\l=-(N_f-N_c)Q{\d z\over z}$
near $P_-$, we obtain
$$
\eqalign{
\R_{P_+}(z^{-1}\d\l)= & 2^{1/N_c}S_2^+,\ \cr
 \R_{P_-}(z^{-1}\d\l)
= & -({\L ^2 \over 2})^{1/(N_f-N_c)}\big(S_2^-+{1\over 2}(1-{1\over N_f-N_c})
t_1^2\big) \cr}
\eqno(3.9)
$$
Subsituting in the values of $\+$ and $\-$
given in (2.2), and rewriting the result in terms
of $\tilde s_2$ and the operator ${\cal D}$ of (1.2),
we can rewrite the renormalization group equation (3.1) as
$$
\eqalign{
2\pi i{\cal D}\F
=- & (N_f-2N_c)
\bigl \{ \tilde s_2-\delta_{N_f,2N_c-2}{\L ^2 \over 4}
-\delta_{N_f,2N_c-1}{\L ^2 \over 4}t_1\bigr \} \cr
& +
(N_f-N_c)t_2-{1\over 2}(N_f-N_c-1)t_1^2
\cr}
\eqno(3.10)
$$
\bigskip
Before proceeding further, we would like to
note a few features of the renormalization group
equation and of our choice of prepotential.

\item{(1)}
The RG equations (3.1) and (3.10)
are actually invariant under a change of cuts. Indeed,
a change of cuts would shift the values of the regularized
integrals (1.4) by a linear expression,
and hence $\F$ by a quadratic expression in the masses $m_j$,
independent of the $a_k$. In view of
Euler's relation, such terms cancel in the left hand side of (3.1)
and (3.10). Thus the right hand side of the RG only transforms
under a change of homology basis, and is a modular form;

\item{(2)} From the point of view
of gauge theories alone, we can in practice ignore
on the right hand side of (3.1) and (3.10) 
terms which do not depend on the $a_k$.
Such terms can always be cancelled by a suitable
$a_k$-independent correction to $\F$. These corrections do not
affect the Wilson effective action since it
depends only on the derivatives of $\F$ with respect to $a_k$;

\item{(3)} Some caution may be necessary in interpreting
$\tilde s_2$, in terms of the classical order parameters
$\bar a_k$. In particular, when $N_f\geq N_c$, there are
several natural ways of parametrizing the curve (2.3),
which the $\a_k$ get shifted in different ways
to $\tilde a_k\not= a_k$ [3][4]. As noted in [6],
the prepotential $\F$ is independent of such redefinitions
of the $\bar a_k$. However, this would
of course not be the case
for $\bar s_2\equiv\sum_{k<j}^{N_c}\bar a_k\bar a_j$,
which argues for a distinct interpretation for
$\tilde s_2=\sum_{j<k}\tilde a_k\tilde a_j$.
\bigskip
\noindent{\bf (c) The renormalization group equation 
in terms of $\vartheta$-functions}
\medskip
As noted above, the right hand side of the RG equation
(3.1) is in general a modular form. For $N_f=0$ (and arbitrary
$N_c$), we can exploit the symmetry between the branch points
$x_k^{\pm}$ given by
$y^2=(A-\L)(A+\L)=\prod_{k=1}^{N_c}(Q-x_k^+)(Q-x_k^-)$ 
and known formulae for their cross-ratios to write it
explicitly in terms of $\vartheta$-functions. More precisely,
we observe that
$$
\sum_{k=1}^{N_c}\a_k^2=\sum_{k=1}^{N_c}(x_k^+)^2
=\sum_{k=1}^{N_c}(x_k^+)^2\eqno(3.11)
$$
Let the canonical homology basis be given by $A_k$ cycles
surrounding the cut from $x_k^-$ to $x_k^+$, $1\leq k\leq N_c-1$
on one sheet, and by $B_k$ cycles going
from $x_{N_c}^-$ to $x_k^-$ on one sheet, and coming
back from $x_k^-$ to $x_{N_c}^-$ on the opposite sheet.
Then for the dual basis of Abelian differentials
$\d\o=(\d\o_k)_{k=1,\cdots,N_c-1}$, we introduce the
basis vectors $e^{(k)}$ and $\tau^{(k)}$ of the Jacobian
lattice by
$$
\oint_{A_k}\d\o=e^{(k)},\ \ \oint_{B_k}\d\o=\tau^{(k)}
$$
We have then the following relations between points in the
Jacobian lattice
$$
\int_{x_k^-}^{x_k^+}\d\o={1\over 2}e^{(k)},
\ \ 
\int_{x_k^+}^{x_{k+1}^-}\d\o={1\over 2}(\tau^{(k+1)}
+\tau^{(k)})\eqno(3.12)
$$
Let $\phi(Q)$ denote the Abel map
$$
\phi(P)=(\int_{Q_0}^Q\d\o_1,\cdots,\int_{Q_0}^Q\d\o_{N_c-1})
$$
If we choose $Q_0$ so that $\phi(x_1^-)={1 \over 2} \tau ^{(1)}$, 
it follows from (3.12) that
$$
\eqalignno{
\phi(x_k^-)=&{1\over 2}(e^{(1)}+\cdots+e^{(k-1)})+{1\over 2}\tau^{(k)},\ 
1\leq k\leq N_c-1\cr
\phi(x_k^+)=&{1\over 2}(e^{(1)}+\cdots+e^{(k)})+{1\over 2}\tau^{(k)},\ 
1\leq k\leq N_c-1\cr
\phi(x_{N_c}^-)=&{1\over 2}(e^{(1)}+\cdots+e^{(N_c-1)}\cr
\phi(x_{N_c}^+)=&0&(3.13)\cr}
$$
If we introduce the functions $F_l^k(Q)$ by
$$
F_l^k(Q)={\vartheta(\phi(x_l^-+x_k^+ +Q)|\tau) ^2
\over 
\vartheta(\phi(x_{N_c}^-+x_k^+ +Q)|\tau) ^2}\eqno(3.14)
$$
an inspection of the zeroes shows that we have the following relation
between $F_l^k$ and cross-ratios
$$
{F_l^k(Q')\over
F_l^k(Q)}
={Q'-x_l^-\over Q-x_l^-}{Q-x_{N_c}^-\over Q'-x_{N_c}^-}\eqno(3.15)
$$
For the Riemann surface (2.2), we also have for all $Q$
$$
\prod_{l=1}^{N_c}(Q-x_l^+)=A(Q)-\L=\prod_{l=1}^{N_c}(Q-x_l^-)-2\L
\eqno(3.16)
$$
Setting $Q=x_k^+$ gives the relation 
$$
\prod_{l=1}^{N_c}(x_k^+-x_l^-)=2\L\eqno(3.17)
$$
Combining with products of expressions of the form (3.15)
evaluated at branch points, we can actually identify the branch
points
$$
\eqalignno{
x_k^+-x_{N_c}^-=&\Lambda G_k\cr
x_k^+-x_l^+=&\Lambda (G_k-G_l)\cr
x_k^+=&-{\Lambda \over N_c}\sum_{l=1}^{N_c}G_l+\Lambda G_k&(3.18)\cr}
$$
where $G_k$ is defined to be
$$
G_k=2^{1\over N_c}\prod_{l=1}^{N_c} \biggl \{
\big[{F_l^k(x_m^-)\over F_l^k(x_k^+)}\big]^{1\over N_c}
\prod_{k'=1}^{N_c}
\big[{F_l^{k'}(x_{k'}^+)\over F_l^{k'}(x_m^-)}\big]^{1\over N_c ^2 } \biggr \}
\eqno(3.19)
$$
Since $F_l^k (x_k^-)$ is independent of $k$, this expression may be simplified,
$$
G_k=2^{1\over N_c}\prod_{l=1}^{N_c} 
\prod_{k'=1}^{N_c}
\bigg [ {F_l^{k}(x_m^-)\over F_l^{k'}(x_m^-)}\bigg ]^{1\over N_c ^2 } 
\eqno(3.20)
$$
The evaluation of the functions $F_l^k$ on the branch points is 
particularly simple,
and we have
$$
F_l^k(x_m^-) 
= { \vartheta ( \phi (x_l ^-) + \phi (x_m ^-) + \phi (x_k ^+) | \tau )^2
\over 
\vartheta ( \phi (x_{N_c} ^-) + \phi (x_m ^-) + \phi (x_k ^+) | \tau )^2}
\eqno (3.21)
$$
where the values of $\phi (x ^\pm)$ can be read off from (3.13).
 This leads to the following expression for the right hand side of
(3.10) :
$$
\sum_{k=1}^{N_c}\tilde a_k^2=\Lambda ^2 \sum_{k=1}^{N_c}Q_k^2-
{\Lambda ^2\over N_c}\big(\sum_{k=1}^{N_c}Q_k\big)^2
\eqno(3.22)
$$
which is a modular form.

\bigskip
\bigskip

\centerline{\bf IV. THE WEAK-COUPLING LIMIT}

\bigskip

It is instructive to verify the renormalization group equation (3.10) in 
the weak-coupling limit analyzed in [6] to 2-instanton order.
\medskip
We recall the expression obtained in [6] for
the prepotential $\F$ to two-instanton order
in the regime of $\Lambda\rightarrow 0$. Let the functions
$S_(x)$ and $S_k(x)$ be defined by
$$
S(x)={\prod_{j=1}^{N_f}(x+m_j)\over\prod_{l=1}^{N_c}(x-a_l)^2}
={S_k(x)\over (x-a_k)^2}
\eqno(4.1)
$$
Then the prepotential $\F$ is given by
$$
\F=\F^{(0)}+\F^{(1)}+\F^{(2)}+O(\L^6)
$$
with the terms $\F^{(0)}$, $\F^{(1)}$, $\F^{(2)}$ corresponding respectively
to the one-loop perturbative
contribution, the 1-instanton contribution, and the 2-instanton contribution
$$
\eqalignno{
2\pi i\F^{(0)}=&-{1\over 4}\sum(a_k-a_l)^2\ln {(a_k-a_l)^2\over\Lambda^2}
+{1\over 4}\sum_{j,k}(a_k+m_j)^2\ln {(a_k+m_j)^2\over\Lambda^2}\cr
2\pi i\F^{(1)}=&{1\over 4}\L^2\sum_{k=1}^{N_c}S_k(a_k)\cr
2\pi i\F^{(2)}=&{1\over 16}\L^4\big
(\sum_{k\not=l}{S_k(a_k)S_l(a_l)\over(a_k-a_l)^2}
+{1\over 4}\sum_{k=1}^{N_c}S_k(a_k)\D_{a_k}^2S_k(a_k)\big)&(4.2)\cr}
$$
Here we have ignored quadratic terms in $a_k$, since they are automatically
annihilated by the operator ${\cal D}$. We also note that
the arguments of [6] only determine $\F$ up to $a_k$-independent terms,
and thus we shall drop all such terms in the subsequent
considerations.
The formulae (4.2) imply
$$
\sum_{k=1}^{N_c}a_k{\D\F\over\D a_k}
+\sum_{j=1}^{N_f}m_j{\D\F\over\D m_j}-2\F
=(N_f-2N_c)\big({1\over 4\pi i}\sum_{k=1}^{N_c}a_k^2
+\F^{(1)}+2\F^{(2)}\big)\eqno(4.3)
$$
where all $\L^6$ terms have been ignored.

On the other hand, up to $a_k$-independent terms,
the renormalization group equation (3.10) reads
$$
\sum_{k=1}^{N_c}a_k{\D\F\over\D a_k}
+\sum_{j=1}^{N_f}m_j{\D\F\over\D m_j}-2\F
={1\over 4\pi i}(N_f-2N_c)\sum_{k=1}^{N_c}\tilde a_k^2
\eqno(4.4)
$$
where we have rewritten $\tilde s_2$ as
$$
\tilde s_2=-{1\over 2}\sum_{k=1}^{N_c}\tilde a_k^2+
{\L^2\over 16}\delta_{N_f,2N_c-1}\eqno(4.5)
$$ 
To compare (4.3) with (4.4) 
we need first to evaluate $\sum_{k=1}^{N_c}\tilde a_k^2$ in terms of
the renormalized order parameters $a_k$. 
Using the formula (3.11) of [6], this can be done routinely 
$$
a_k=\a_k+{\L^2\over 4}\tD_k\S_k(\a_k)+{\L^4\over 64}\tD_k^3\S_k(\a_k)
+O(\L^6)\eqno(4.6)
$$
where we have set $\tD_k=\D/\D\a_k$, and
defined functions $\S(x)$, $\S_k(x)$ in analogy
with (4.1), but with $a_k$ replaced by $\a_k$.
Inverting $\a_k$ in terms of $a_k$, and rewriting the result in terms
of the derivatives $\D_k=\D/\D a_k$ with respect to the renormalized
parameters $a_k$, we find
$$
\a_k=a_k-{\L^2\over 4}\D_kS_k(a_k)-{\L^4\over 64}\D_k^3S_k(a_k)^2
+{\L^4\over 16}\sum_{l=1}^{N_c}\D_lS_l(a_l)\D_k\D_lS_k(a_l)+O(\L^6)
\eqno(4.7)
$$
and hence
$$
\eqalignno{
\sum_{k=1}^{N_c}\tilde a_2^2
=&\sum_{k=1}^{N_c}a_k^2-{\L^2\over 2}\sum_{k=1}^{N_c}a_k\D_kS_k(a_k)
-{\L^4\over 32}\sum_{k=1}^{N_c}a_k\D_k^3S_k(a_k)^2\cr
&+{\L^4\over 8}\sum_{k,l=1}^{N_c}a_k\D_lS_l(a_l)\D_k\D_lS_k(a_k)
+{\L^4\over 16}\sum_{k=1}^{N_c}(\D_kS_k(a_k))^2+O(\L^6)
&(4.8)\cr}
$$
Next, we need a number of identities
which can be established by contour integrals, in analogy with the
identities in Appendix B of [6]
$$
\eqalignno{
\sum_{k=1}^{N_c}a_k\D_kS_k(a_k)=&-\sum_{k=1}^{N_c}S_k(a_k)+
\{a_k{\rm -independent\ terms}\}\cr
\sum_{k=1}^{N_c}a_k\D_k^3S_k(a_k)^2=&-3\sum_{k=1}^{N_c}\D_k^2S_k(a_k)^2\cr
\sum_{k,l}a_k\D_lS_l(a_l)\D_k\D_lS_k(a_k)
=&-2\sum_{l=1}^{N_c}(\D_lS_l(a_l))^2
+2\sum_{k\not=l}{S_k(a_k)S_l(a_l)\over (a_k-a_l)^2}\cr
&\qquad-\sum_{k\not=l}S_k(a_k)\D_k^2S_k(a_k)&(4.9)\cr}
$$
Using (4.9) we can indeed recast $\sum_{k=1}^{N_c}\tilde a_k^2$ as
$$
\sum_{k=1}^{N_c}
\tilde a_k^2=
\sum_{k=1}^{N_c}a_k^2+\sum_{k=1}^{N_c}
{\L^2\over 2}S_k(a_k)
+{\L^4\over 4}\big(\sum_{k\not=l}
{S_k(a_k)S_l(a_l)\over (a_k-a_l)^2}
+{1\over 4}\sum_{k=1}^{N_c}
S_k(a_k)\D_k^2S_k(a_k)\big)
\eqno(4.10)
$$
The equality of the two right hand sides in
(4.3) and (4.4) follows.
\vfill\break

\centerline{\bf REFERENCES}
\bigskip

\item{[1]} N. Seiberg and E. Witten, Nucl. Phys. B 426 (1994) 19, 
hep-th/9407087;
Nucl. Phys. B 431 (1994) 484, hep-th/9408099.
\item{[2]} A. Klemm, W. Lerche, S. Yankielowicz, and S. Theisen,
Phys. Lett. B 344 (1995) 169;\hfil\break
P.C. Argyres and A. Faraggi, Phys. Rev. Lett. 73 (1995) 3931;
\hfil\break
P.C. Argyres, R. Plesser, and A. Shapere, Phys. Rev. Lett. 75 (1995) 1699,
hep-th/9505100;\hfil\break
J. Minahan and D. Nemeshansky, hep-th/9507032;\hfil\break
U.H. Danielsson and B. Sundborg, Phys. Lett. B 358 (1995) 273,
USITP-95-12, UUITP-20/95;
hep-th/9504102;\hfil\break
A. Brandhuber and K. Landsteiner, Phys. Lett. B 358 (1995) 73, 
hep-th/9507008;
\hfil\break
M. Alishahiha, F. Ardalan, and F. Mansouri, hep-th/9512005.
\item{[3]} A. Hanany and Y. Oz, Nucl. Phys. B 452 (1995) 73, 
hep-th/9505075\hfil\break
P.C. Argyres and A. Shapere, hep-th/9609175;\hfil\break
A. Hanany, hep-th/9509176.
\item{[4]} I.M. Krichever and D.H. Phong, hep-th/9604199, 
J. of Differential Geometry;
\item{[5]} R. Donagi and E. Witten, hep-th/9511101;
\item{[6]} E. D'Hoker, I.M. Krichever, and D.H. Phong, hep-th/9609041;
hep-th/9609145;
\item{[7]} S. Kachru, A. Klemm, W. Lerche, P. Mayr, and C. Vafa,
Nucl. Phys. B 459 (1996) 537, hep-th/9508155;
\item{[8]} M. Matone, Phys. Lett. B 357 (1995) 342;\hfil\break
G. Bonelli and M. Matone, Phys. Rev. Lett. 76 (1996) 4107;\hfil\break
G. Bonelli and M. Matone, hep-th/9605090;
\item{[9]} T. Eguchi and S.K. Yang, Mod. Phys. Lett. A 11 (1996) 131;
\hfil\break
J. Sonnenschein, S. Theisen, and S. Yankielowicz, hep-th/9510129.
\item{[10]} I.M. Krichever, Comm. Pure Appl. Math. 47 (1994) 437.
\item{[11]} A. Gorsky, I.M. Krichever, A. Marshakov, A. Mironov, 
and A. Morozov,
Phys. Lett. B 355 (1995) 466, hep-th/9505035;\hfil\break
H. Itoyama and A. Morozov, hep-th/9512161, hep-th/9601168;\hfil\break
A. Marshakov, hep-th/9602005\hfil\break 
E. Martinec and N. Warner, hep-th/9509161, hep-th/9511052;\hfil\break
E. Martinec, hep-th/9510204,\hfil\break
C. Anh and S. Nam, hep-th/9603028,\hfil\break
T. Nakatsu and K. Takasaki, hep-th/9509162;
\item{[12]} A. Klemm, W. Lerche, and S. Theisen, hep-th/9505150;\hfil\break
K. Ito and S.K. Yang, hep-th/96
\item{[13]} K. Ito and N. Sasakura, SLAC-PUB-KEK-TH-470, hep-th/9602073;
\item{[14]} N. Dorey, V. Khoze, and M. Mattis, hep-th/9606199, 
hep-th/9607202;\hfil\break
Y. Ohta, hep-th/9604051, hep-th/9604059;\hfil\break
A. Klemm, W. Lerche, P. Mayr, C. Vafa, and N. Warner, hep-th/9604034;
\item{[15]} D. Finnell and P. Pouliot, Nucl. Phys. {\bf B453} (1995) 225;
\hfil\break
A. Yung, hep-th/9605096; \hfil\break
F. Fucito and G. Travaglini, hep-th/9605215.
\item{[16]} T. Harano and M. Sato, hep-th/9608060
 
\end